\documentclass{Interspeech2024}
\usepackage{multirow}
\usepackage{footnote}
\usepackage{footmisc}
\usepackage{hyperref}
\usepackage[T1]{fontenc}
\makesavenoteenv{tabular}
\makesavenoteenv{table}

\renewcommand{\thefootnote}{\alph{footnote}}

\newcommand{\astfootnote}[1]{%
\let\oldthefootnote=\thefootnote%
\setcounter{footnote}{0}%
\renewcommand{\thefootnote}{\fnsymbol{footnote}}%
\footnote{#1}%
\let\thefootnote=\oldthefootnote%
}

\interspeechcameraready

\title{Temporal-Channel Modeling in Multi-head Self-Attention\\ for Synthetic Speech Detection\thanks{The corresponding author is Ruijie Tao}}

\vspace{-1cm}
\name[affiliation={1}]{Duc-Tuan}{Truong}
\name[affiliation={2}]{Ruijie}{Tao}
\name[affiliation={3}]{Tuan}{Nguyen}
\name[affiliation={1}]{Hieu-Thi}{Luong}
\name[affiliation={4}]{Kong Aik}{Lee}
\name[affiliation={1}]{Eng Siong}{Chng}

\address{
  $^1$Nanyang Technological University, Singapore \quad
  $^2$National University of Singapore, Singapore \\
  $^3$Institute for Infocomm Research (I$^{2}$R), A$^\star$STAR, Singapore \\
  $^4$The Hong Kong Polytechnic University, Hong Kong}

\email{truongdu001@e.ntu.edu.sg, ruijie@nus.edu.sg, stunvat@i2r.a-star.edu.sg, hieuthi.luong@ntu.edu.sg, kong-aik.lee@polyu.edu.hk, aseschng@ntu.edu.sg}


\begin{document}

\maketitle


\begin{abstract}
Recent synthetic speech detectors leveraging the Transformer model have superior performance compared to the convolutional neural network counterparts. This improvement could be due to the powerful modeling ability of the multi-head self-attention (MHSA) in the Transformer model, which learns the temporal relationship of each input token. However, artifacts of synthetic speech can be located in specific regions of both frequency channels and temporal segments, while MHSA neglects this temporal-channel dependency of the input sequence. In this work, we proposed a Temporal-Channel Modeling (TCM) module to enhance MHSA's capability for capturing temporal-channel dependencies. Experimental results on the ASVspoof 2021 show that with only 0.03M additional parameters, the TCM module can outperform the state-of-the-art system by 9.25\% in EER. Further ablation study reveals that utilizing both temporal and channel information yields the most improvement for detecting synthetic speech\footnote{Code and pre-trained models are available at \href{https://github.com/ductuantruong/tcm_add}{https://github.com/ductuantruong/tcm\_add}}.


    
\end{abstract}

\section{Introduction}

Powered by advanced deep generative neural networks, recent text-to-speech (TTS) and voice conversion (VC) systems have the ability to generate highly realistic synthetic human voices. Although this application can benefit many areas including data augmentation \cite{deepfake_aug}, criminals can utilize these fake speeches for malicious purposes leading to financial fraud, political conflict, and impersonation. Due to that, synthetic speech detection has been an active research field \cite{sdd_servey_1,sdd_servey_2}. To capture local synthetic artifacts, convolutional neural networks (CNNs)  have conventionally served as the foundational architecture for SSD models. This approach covers a wide range of CNNs including LCNNs \cite{lightcnn1, lightcnn2}, residual-connected ResNet \cite{resnet2, resnet3}, and other variants \cite{othercnn3, othercnn4}. However, CNN-based models exhibit limitations in capturing the long-range dependencies of the input sequence. To overcome this, numerous studies employ Transformer models \cite{transformer1,transformer2,mfaconformer}, yielding improved performance over CNN-based SSD models. 



Notably, the recent SSD model \cite{conformer}, which combines the rich sequence representation of a self-supervised learning (SSL) model XLSR and the transformer-based Conformer architecture, achieves the state-of-the-art result in the ASVspoof 2021 corpus. This improvement can be attributed to the powerful modeling capability of the multi-head self-attention (MHSA) mechanism. It is conjecture that artifact details of synthetic speech can be located in specific regions of both temporal and spectral domain \cite{exp_tcd_1, exp_tcd_2, exp_tcd_3}. Therefore, incorporating the relationship between temporal and spectral information can provide a more complete and accurate representation for detecting artifacts in synthetic speech. By leveraging the temporal and spectral dependencies, several SSD systems \cite{graph2,graph4} exhibit improved capabilities in detecting deepfake speech. However, the MHSA in transformer-based SSD systems focuses on computing dot product between input tokens along the temporal dimension, hence it may overlook the dependencies between the temporal and channel dimensions of input sequences, which can be crucial for SSD tasks.
 

To better leverage the temporal and channel interaction of the input sequence for the XLSR-Conformer system, we propose the Temporal-Channel Modeling (TCM) module in the multi-head self-attention of the Conformer model. Our TCM module is based on the \textit{head tokens} design, in which each head token represents the information on the channel dimension. The idea of head tokens is first proposed in \cite{headtoken} to enhance the interaction between the representation of attention heads in the MHSA and have improved the performance of Vision Transformer trained in the small-scale image classification dataset. However, in this work, head tokens aim to facilitate the correlation between temporal and channel dependencies by interacting them with the temporal tokens during MHSA. We also modify the original head token design by enriching the classification token with both temporal and channel information. The proposed TCM module, wherein with a marginal increase in parameters, improves the performance of the state-of-the-art XLSR-Conformer system on the ASV2021 eval set. Through empirical evaluation of the contribution of each component in the TCM module, temporal information from input tokens and channel information from head tokens both play an important role in the improvement of the TCM module.

\begin{figure*}[t]
  \centering
  \includegraphics[width=1.0\textwidth]{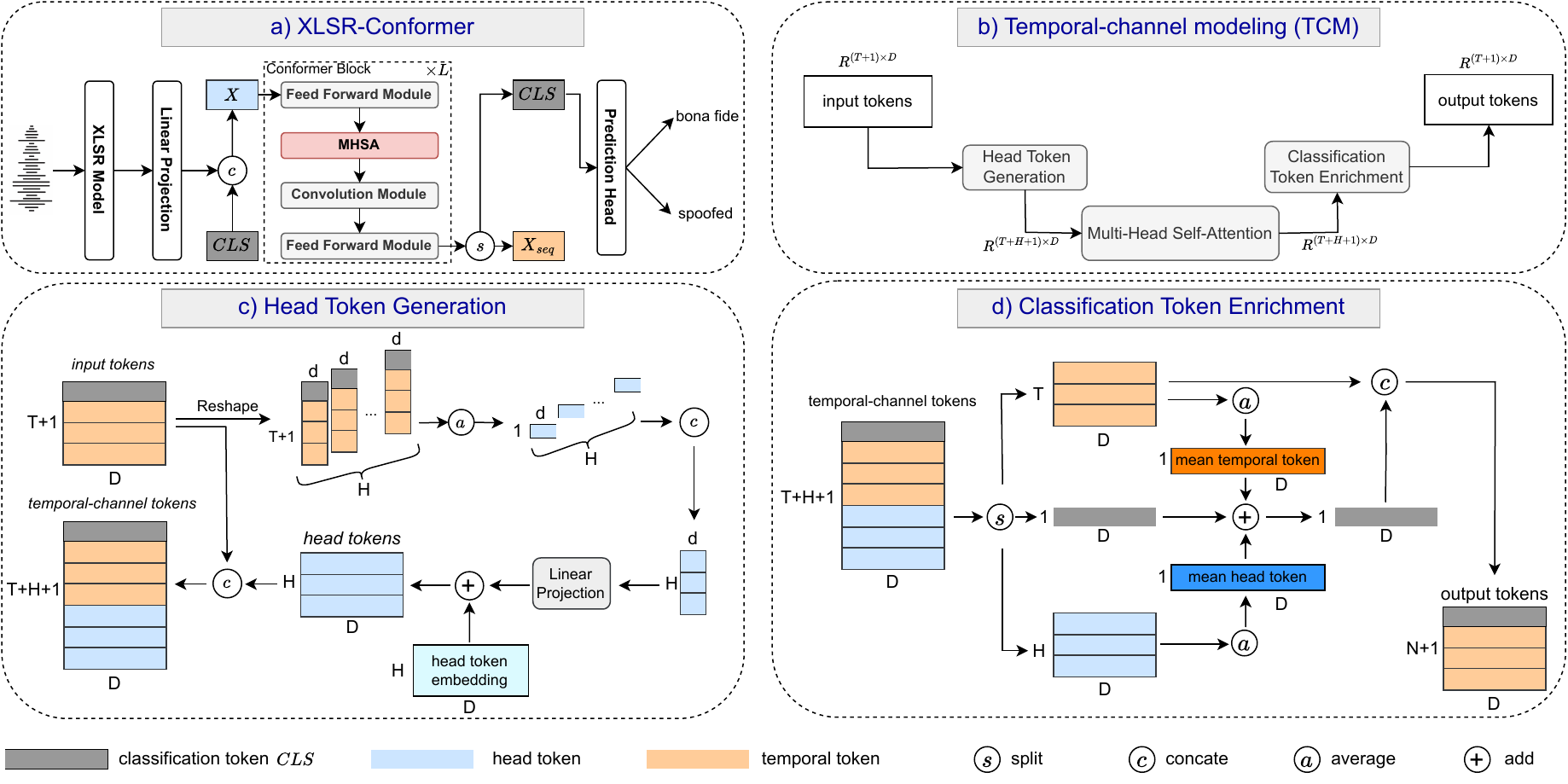}
  \vspace{2mm}
  \caption{The overall architecture of the baseline XLSR-Conformer and our proposed temporal-channel modeling (TCM) module. The TCM module is used to replace the multi-head self-attention (MHSA) of each Conformer block in the baseline XLSR-Conformer. The TCM module architecture includes three main parts: Head Token Generation, Multi-Head Self-Attention, and Classification Token Enrichment. The objective of TCM is to generate the head token for channel information and then integrate the temporal-channel dependency into the original temporal tokens for better synthetic speech detection. }
  \label{fig:tcm_arch}
\end{figure*}
\section{Method}
\subsection{Baseline XLSR-Conformer}
We adopt the state-of-the-art XLSR-Conformer \cite{conformer} as our baseline architecture. As illustrated in Figure~\ref{fig:tcm_arch}.a, it leverages the pre-trained XLSR \cite{xls_r}, a variant of the wav2vec 2.0 model. Benefiting from the large-scale architecture and training on extensive data in an SSL manner, SSL models including XLSR can extract rich speech representations that have been useful for numerous speech tasks \cite{ssl_app_asr,ssl_app_sv,ssl_app_emotion,ssl_app_age_1,ssl_app_age_2} including synthetic speech detection \cite{graph3}. XLSR comprises two main components: a CNN front-end to transfer the 1D raw waveform into 2D temporal-channel representation, and $24$ transformer encoder layers for capturing the global relationship of the speech. The shape of the output speech representation is $(T \times D)$, where $T$ denotes the temporal length and $D$ is the channel dimension of XLSR representation.

After that, the XLSR representation is projected to $D$-dimensional and concatenated with the learnable classification token $CLS$ to form an input sequence $X \in \mathbb{R}^{(T+1) \times D}$ for the Conformer model. The Conformer model consists of $L$ Conformer blocks, each Conformer block includes the MHSA, feed-forward module, and the additional Convolutional layer to capture local dependencies within the speech representation. Finally, the $CLS$ token is detached from the Conformer model's output representation to determine whether the input speech is bona fide or spoof.
\vspace{2mm}
\subsection{Temporal-Channel Modeling module}
The study of \cite{headtoken} introduces the concept of head token design, which initially focuses on fostering interaction between attention heads in multi-head self-attention (MHSA) and has improved the performance of image classification models trained on limited datasets. While the Temporal-Channel Modeling (TCM) approach is inspired by the head token design, its goal is to assist multi-head self-attention in capturing temporal-channel dependencies which can be essential for detecting synthetic speech. The proposed TCM module replaces the original MHSA of each Conformer block in the baseline model. As shown in Figure \ref{fig:tcm_arch}.b, the TCM architecture comprises three parts: Head Token Generation, Multi-Head Self-Attention, and Classification Token Enrichment. Similar to MHSA, TCM will not change the shape of the input and output token sequences for each Conformer block. 

\subsubsection{Head Token Generation}
The Temporal-Channel Modeling module begins with the Head Token Generation component, designed to generate head tokens that represent the channel information of the input. These tokens interact with temporal information in subsequent steps. As shown in Figure~\ref{fig:tcm_arch}.c, the input sequence of the Head Token Generation component consists of classification token $CLS$ and temporal tokens $X \in \mathbb{R}^{(T+1) \times D}$. It first undergoes the head token generation process where $X$ is reshaped into $H$ segments of $d = D / H$ dimensions along the channel axis, where $H$ is the number of attention heads in MHSA. Subsequently, each segment undergoes temporal average pooling and concatenates together, followed by the linear projection consisting of a fully connected layer and the GeLU function to project back to $D$-dimension channel representation. Since these steps are similar to the MHSA transformation process, by projecting the input sequence into distinct attention heads, these embeddings are designated as head tokens, representing different parts of the channel dimension. To distinguish head tokens from input tokens, we add a learnable head token embedding with the shape of $(H \times D)$ to head tokens. After obtaining head tokens, they are concatenated with the input sequence along the temporal dimension, forming a new temporal-channel token sequence with the length of $(T+H+1)$ to the MHSA.

\subsubsection{Multi-Head Self-Attention}
The multi-head self-attention mechanism within our TCM operates similarly to conventional multi-head self-attention but with the input sequence containing both temporal and channel tokens, rather than just temporal tokens. To learn the temporal-channel interaction for spoofing detection, the multi-head self-attention mechanism transforms the temporal-channel tokens into query $Q$, key $K$, and value $V$. This is achieved by projecting temporal-channel tokens $H$ times using corresponding linear projection matrices $W_i^Q$, $W_i^K$, $W_i^V$, resulting in $d$-dimensional channel representations, where $i$ represents the index of the head within the MHSA. With the scaled dot product, the self-attention operator then calculates appropriate weights for each token along the temporal axis based on its relevance to each other, and this process is repeated in parallel across $H$ attention heads. Subsequently, the output of each head is concatenated and subjected to a final linear projection denoted as $W^O$, yielding the output embedding. The multi-head self-attention can be represented by the following equation:
\begin{align} \label{eq:attention}   
    \text{MultiHead}(X) &= \text{Concat}(\text{head}_{1}, \ldots, \text{head}_{H})W^O \nonumber\\
    \text{where } \text{head}_i &= \text{softmax}(\frac{XW_i^Q \cdot (XW_i^K)^T}{\sqrt{d}}) \cdot XW_i^V
\end{align}
Given that the self-attention of each head is independently computed along the temporal axis of the input sequence, if the input sequence only contains temporal tokens, the model may lack the interaction between the temporal and channel dimensions in the MHSA. However, in the proposed method, head tokens representing channel information are put together with temporal tokens, hence MHSA can learn temporal-channel dependencies by attending to different parts including temporal and head tokens of the input sequence.



\subsubsection{Classification Token Enrichment}
Although the classification token $CLS$ can attend the information from both temporal and channel tokens during MHSA, we further enrich the $CLS$ token with both temporal and channel tokens with the Classification Token Enrichment component module because the $CLS$ token is directly used for the final prediction, and the information from both tokens can be both crucial for detecting artifacts. Figure~\ref{fig:tcm_arch}.d illustrates the Classification Token Enrichment component of the proposed TCM module. Firstly, the temporal and head tokens are segregated from the MHSA output and subjected to average pooling to get the mean temporal token and mean head token. After that, instead of considering only the mean head token in the original head token design \cite{headtoken}, our TCM module also enriches the classification token $CLS$ with the mean temporal token. Finally, the enriched classification token is concatenated with the temporal tokens to form the output sequence, keeping the same shape of $(T+1) \times D$ as the input sequence.
\vspace{2mm}
\section{Experiments}
\begin{table*}[t]
\centering
\begin{tabular}{l|c|cccc|cc}
\hline
\multicolumn{1}{c|}{\multirow{2}{*}{\textbf{System}}} & \multirow{2}{*}{\textbf{Params (M)}} & \multicolumn{2}{c}{\textbf{LA (Fix)}} & \multicolumn{2}{c|}{\textbf{LA (Var)}} & \textbf{DF (Fix)} & \textbf{DF (Var)} \\ \cline{3-8} 
\multicolumn{1}{c|}{}                                 &                                      & EER (\%)         & min t-DCF          & EER (\%)          & min t-DCF          & EER (\%)          & EER (\%)          \\ \hline
RawNet2 \cite{resnet1}                                               & 25.43                                & 9.50             & 0.4257             & -                 & -                  & 22.38             & -                 \\
AASIST \cite{graph2}                                               & 0.30                                 & 5.59             & 0.3398             & -                 & -                  & -                 & -                 \\
RawFormer \cite{transformer2}                                             & 0.37                                 & 4.98             & 0.3186             & 4.53              & 0.3088             & -                 & -                 \\
XLSR-AASIST \cite{graph3}                                          & 317.84                               & \textbf{1.00}       & \textbf{0.2120}    & -                 & -                  & 3.69              & -                 \\
XLSR-Conformer \cite{conformer}                                        & 319.74                               & 1.38             & 0.2216             & \textbf{0.97}     & \textbf{0.2116}    & \underline{2.27}        & \underline{2.58}        \\
XLSR-Conformer (reproduce)                            & 319.74                               & 1.40             & 0.2226             & 1.26              & 0.2200             & 2.79              & 2.98              \\ \hline
XLSR-Conformer + TCM                                  & 319.77                               & \underline{1.03}    & \underline{0.2130}       & \underline{1.18}        & \underline{0.2172}       & \textbf{2.06}     & \textbf{2.25}     \\ \hline
\end{tabular}
\caption{Performance comparison with the state-of-the-art systems on the ASVspoof 2021 eval set with fixed-length (Fix) and variable-length (Var) utterance evaluation (Bold denotes the best result, underline denotes the second-best result, and dash denotes the results are unavailable).}
\label{tab:comparision}
\end{table*}

\subsection{Dataset and metrics}

While the training and development data are from the ASVspoof 2019 \cite{asvspoof2019} logical access (LA) track containing clean speech with text-to-speech and voice conversion attacks, we evaluated our method on the ASVspoof 2021 \cite{asvspoof2021} logical access (LA) and deep fake (DF) tasks. ASVspoof 2021 LA eval set includes 2 known and 11 unknown and the speech data is distorted by various codec and compression variations, mimicking real-world scenarios. Additionally, ASVspoof 2021 introduced a new DF eval consisting of two new additional sets of source data compared to the LA set. Our primary evaluation metrics are the common-used equal error rate (EER) \cite{eer} and minimum normalized tandem detection cost function (t-DCF).
\subsection{Implementation details}
In the training step, the audio data are cropped or concatenated giving segments of approximately 4 seconds duration (64,600 samples). We used the Adam optimizer with a learning rate of $10^{-6}$ with a weight decay of $10^{-4}$ to optimize a weighted cross-entropy loss. The batch size for the training step is set to 20. The final result is reported using the model checkpoint created by averaging the top-5 best validation performance models. Early stopping is applied when the cross entropy loss in the validation set did not improve for 7 epochs. All of the experiments are trained with one Nvidia A40 GPU with the same random seed. In terms of model architecture, following our baseline \cite{conformer}, the pre-trained SSL model XLSR\footnote{\href{https://github.com/pytorch/fairseq/tree/main/examples/wav2vec}{https://github.com/pytorch/fairseq/tree/main/examples/wav2vec}} is utilized as an upstream model to extract intermediate representation from the raw input signal. 


To be comparable with \cite{conformer,graph3}, the signal noise injection data augmentation technique RawBoost \cite{rawboost} is utilized in our experiments.  The configuration and parameters of RawBoost used in our experiment are similar to the original paper. Following our baseline system \cite{conformer}, we trained two separate SSD systems with two different Rawboost settings to evaluate on the LA and DF track, respectively. In the LA track, the SSD system is trained with the RawBoost technique combining linear and non-linear convolutive noise and impulsive signal-dependent additive noise strategies. On the other hand, the stationary signal-independent additive, randomly colored noise, is added during the training in the DF track.  


\subsection{Results}
\subsubsection{Comparison with the state-of-the-art systems}
Table~\ref{tab:comparision} compares the performance of the proposed TCM with our reproduced state-of-the-art XLSR-Conformer and other existing competitive systems on the ASVspoof21 LA and DF evaluation set. In the fixed-length input evaluation on the LA track, adding the proposed TCM module can achieve 25\% EER improvement than the baseline XLSR-Conformer for the pooled EER (1.03 \% vs 1.40\%). While XLSR-Conformer with TCM achieved comparable performance to the top-performing LA system XLSR-AASIST \cite{graph3} in the LA track, it attained a new state-of-the-art result of 2.06\% EER in the DF track, surpassing the previous best-reported result of XLSR-Conformer by 9.25\% in the fixed-length input evaluation. Similar gains can be observed in variable-length utterance evaluation. Notably, while achieving noticeable improvement, our TCM module is lightweight since it adds only $0.03 M$ parameters to the XLSR-Conformer system. In the following sections, we conduct experiments on the reproduced XLSR-Conformer for further analysis of the robustness and efficiency of TCM. 


\subsubsection{Transformer and Conformer comparison}

To verify the robustness and effectiveness of the proposed TCM in the SSD task, we replaced the Conformer block with the Transformer one and conducted the study in Table~\ref{tab:transformer}. We notice that TCM can bring relatively stable improvement for both Conformer and Transformer structures. This can indicate that the learned temporal-channel dependency in TCM can be beneficial for detecting spoofed artifacts regardless of the transformer-based architectures. Furthermore, the Conformer-based system yielded superior performance compared to the corresponding Transformer in the baseline setting as well as with the TCM module. These demonstrate that the local information captured by the Convolution module in Conformer is important for the SSD task.

\begin{table}[t]
\centering
\begin{tabular}{lcccc}
\hline
                                         & \multicolumn{2}{c}{\textbf{21LA}}                                             & \multicolumn{2}{c}{\textbf{21DF}}                                             \\ \cline{2-5} 
\multirow{-2}{*}{\textbf{System}}         & Fix                                   & Var                                   & Fix                                   & Var                                   \\ \hline
XLSR-Transformer                         & 1.60          & 1.44          & 2.24          & 2.49          \\
\multicolumn{1}{l}{XLSR-Transformer + TCM} & 1.51          & 1.91 	  & \textbf{2.02} & 2.34          \\ \hline
XLSR-Conformer                           & 1.40          & 1.26          & 2.33          & 2.48                                  \\
XLSR-Conformer + TCM                       & \textbf{1.03} & \textbf{1.18} & 2.06          & \textbf{2.25} \\ \hline
\end{tabular}
\vspace{2mm}
\caption{EER (\%) results to evaluate the robustness of TCM for the Transformer and Conformer Block.}
\label{tab:transformer}
\end{table}
\subsubsection{Multi-head attention}
Table~\ref{tab:number_head} further studies the effect of the different numbers of heads with and without TCM on the ASV2021 LA \& DF eval set. The system with 4 heads can lead to the best performance. (1.03 \% EER on LA track and 2.06 \% EER on DF track.). The improvement by TCM is robust for most cases, except the DF eval track with 8 heads. It is important that an increase in the number of attention heads does not necessarily ensure improved results.
\subsubsection{Ablation study}
Table~\ref{tab:ablation} presents an analysis of the contributions of each component within our TCM module on the ASV2021 DF evaluation set. We observed that the inclusion of head token embeddings leads to a slight improvement in system performance. Conversely, the performance of the TCM module experiences a notable decline, from 2.06\% to 2.40\% EER, when head tokens are excluded from the multi-head attention mechanism or when the mean head token (mean HT) is omitted from addition to the $CLS$ token. A similar trend is observed when the mean temporal token (mean TT) is excluded from enriching the $CLS$ token. Notably, when both mean HT and mean TT are absent, the performance drops significantly to 3.25\% EER, indicating a deterioration compared to the baseline system. These findings underscore the importance of leveraging both temporal and channel information, as represented by temporal tokens and head tokens, in the task of detecting synthetic speech.


\begin{table}[]
\centering
\begin{tabular}{llccc}
\hline
\multirow{2}{*}{\textbf{Track}} & \multicolumn{1}{c}{\multirow{2}{*}{\textbf{System}}} & \multicolumn{3}{c}{\textbf{EER (\%)}} \\ \cline{3-5} 
                                     & \multicolumn{1}{c}{}                                 & H=4         & H=6        & H=8        \\ \hline
\multirow{2}{*}{LA}                  & XLSR-Conformer                                             & 1.40        & 1.14       & 1.72       \\
                                     & XLSR-Conformer + TCM                                       & \textbf{1.03}        & \textbf{1.13}       & \textbf{1.06}       \\ \hline
\multirow{2}{*}{DF}                  & XLSR-Conformer                                             & 2.79        & 2.87       & \textbf{3.11}       \\
                                     & XLSR-Conformer + TCM                                       & \textbf{2.06}        & \textbf{2.84}       & 3.81       \\ \hline
\end{tabular}
\vspace{2mm}
\caption{Our methods with different numbers of heads on ASV2021 LA \& DF eval set.}
\label{tab:number_head}
\end{table}

\begin{table}[t]
\centering
\begin{tabular}{lc}
\hline
                               & \textbf{EER (\%)} \\ \hline
XLSR-Conformer w/o TCM (baseline)           & 2.79 \\           
XLSR-Conformer + TCM           & \textbf{2.06}              \\ \hline
w/o HT embedding               & 2.08              \\
w/o HT in MHSA                 & 2.40              \\
w/o adding mean HT to $CLS$        & 2.41              \\
w/o adding mean TT to $CLS$        & 2.33              \\
w/o adding mean HT \& Mean TT to $CLS$ & 3.25              \\ \hline
\end{tabular}
\vspace{2mm}
\caption{Ablation study of each component in our proposed TCM on ASV2021 DF evaluation set. HT and TT represent head tokens and temporal tokens, respectively.}
\label{tab:ablation}
\end{table}
\section{Conclusion}
In this paper, we propose a Temporal-Channel Modeling module for MHSA-based synthetic speech detection systems. Our method integrates the channel representation head token into the temporal input token within the multi-head self-attention, which forces the model to learn the temporal-channel dependencies from the input sequence. The XLSR-Conformer using our TCM module outperforms the state-of-the-art performance and outperforms competing methods on the ASVspoof 2021 eval set. Additionally, the ablation study validates the effectiveness of our proposed method and demonstrates the importance of temporal-channel modeling in synthetic speech detection.
\section{Acknowledgement}
This research is supported by the National Research Foundation Singapore under the AI Singapore Programme (AISG Award No.: AISG-TC-2023-011-SGIL). Any opinions, findings and conclusions or recommendations expressed in this material are those of the author(s) and do not reflect the views of National Research Foundation, Singapore. The computational work for this article was partially performed on resources of the National Supercomputing Centre, Singapore (https://www.nscc.sg).

\bibliographystyle{IEEEtran}
\bibliography{mybib}

\begin{thebibliography}{10}
\providecommand{\url}[1]{#1}
\csname url@samestyle\endcsname
\providecommand{\newblock}{\relax}
\providecommand{\bibinfo}[2]{#2}
\providecommand{\BIBentrySTDinterwordspacing}{\spaceskip=0pt\relax}
\providecommand{\BIBentryALTinterwordstretchfactor}{4}
\providecommand{\BIBentryALTinterwordspacing}{\spaceskip=\fontdimen2\font plus
\BIBentryALTinterwordstretchfactor\fontdimen3\font minus \fontdimen4\font\relax}
\providecommand{\BIBforeignlanguage}[2]{{%
\expandafter\ifx\csname l@#1\endcsname\relax
\typeout{** WARNING: IEEEtran.bst: No hyphenation pattern has been}%
\typeout{** loaded for the language `#1'. Using the pattern for}%
\typeout{** the default language instead.}%
\else
\language=\csname l@#1\endcsname
\fi
#2}}
\providecommand{\BIBdecl}{\relax}
\BIBdecl

\bibitem{deepfake_aug}
K.~C. Yuen, L.~Haoyang, and C.~E. Siong, ``Asr model adaptation for rare words using synthetic data generated by multiple text-to-speech systems,'' in \emph{2023 Asia Pacific Signal and Information Processing Association Annual Summit and Conference (APSIPA ASC)}, 2023, pp. 1771--1778.

\bibitem{sdd_servey_1}
H.~Wu, J.~Kang, L.~Meng, H.~Meng, and H.~yi~Lee, ``The defender's perspective on automatic speaker verification: An overview,'' 2023.

\bibitem{sdd_servey_2}
A.~Khan, K.~M. Malik, J.~Ryan, and M.~Saravanan, ``Voice spoofing countermeasures: Taxonomy, state-of-the-art, experimental analysis of generalizability, open challenges, and the way forward,'' \emph{arXiv preprint arXiv:2210.00417}, 2022.

\bibitem{lightcnn1}
Z.~Wu, R.~K. Das, J.~Yang, and H.~Li, ``Light convolutional neural network with feature genuinization for detection of synthetic speech attacks,'' in \emph{Proc. INTERSPEECH}, 2020.

\bibitem{lightcnn2}
G.~Lavrentyeva, S.~Novoselov, A.~Tseren, M.~Volkova, A.~Gorlanov, and A.~Kozlov, ``Stc antispoofing systems for the asvspoof2019 challenge,'' in \emph{Proc. INTERSPEECH}, 2019.

\bibitem{resnet2}
X.~Li, N.~Li, C.~Weng, X.~Liu, D.~Su, D.~Yu, and H.~M. Meng, ``Replay and synthetic speech detection with res2net architecture,'' \emph{IEEE International Conference on Acoustics, Speech and Signal Processing (ICASSP)}, pp. 6354--6358, 2021.

\bibitem{resnet3}
X.~Li, X.~Wu, H.~Lu, X.~Liu, and H.~Meng, ``Channel-wise gated res2net: Towards robust detection of synthetic speech attacks,'' \emph{Proc. INTERSPEECH}, 2021.

\bibitem{othercnn3}
N.~Müller, P.~Czempin, F.~Diekmann, A.~Froghyar, and K.~Böttinger, ``{Does Audio Deepfake Detection Generalize?}'' in \emph{Proc. INTERSPEECH}, 2022, pp. 2783--2787.

\bibitem{othercnn4}
A.~M. Rostami, M.~M. Homayounpour, and A.~Nickabadi, ``Efficient attention branch network with combined loss function for automatic speaker verification spoof detection,'' \emph{Circuits, Systems, and Signal Processing}, pp. 1 -- 19, 2021.

\bibitem{transformer1}
C.~Li, F.~Yang, and J.~Yang, ``The role of long-term dependency in synthetic speech detection,'' \emph{IEEE Signal Processing Letters}, vol.~29, pp. 1142--1146, 2022.

\bibitem{transformer2}
X.~Liu, M.~Liu, L.~Wang, K.~A. Lee, H.~Zhang, and J.~Dang, ``Leveraging positional-related local-global dependency for synthetic speech detection,'' in \emph{IEEE International Conference on Acoustics, Speech and Signal Processing (ICASSP)}, 2023, pp. 1--5.

\bibitem{mfaconformer}
H.~seo Shin, J.~Heo, J.~ho~Kim, C.~yeong Lim, W.~Kim, and H.-J. Yu, ``Hm-conformer: A conformer-based audio deepfake detection system with hierarchical pooling and multi-level classification token aggregation methods,'' 2023.

\bibitem{conformer}
E.~Rosello, A.~Gomez-Alanis, A.~M. Gomez, and A.~Peinado, ``{A conformer-based classifier for variable-length utterance processing in anti-spoofing},'' in \emph{Proc. INTERSPEECH}, 2023, pp. 5281--5285.

\bibitem{exp_tcd_1}
J.~Yang, R.~K. Das, and H.~Li, ``Significance of subband features for synthetic speech detection,'' \emph{IEEE Transactions on Information Forensics and Security}, vol.~15, pp. 2160--2170, 2020.

\bibitem{exp_tcd_2}
K.~Sriskandaraja, V.~Sethu, P.~N. Le, and E.~Ambikairajah, ``Investigation of sub-band discriminative information between spoofed and genuine speech,'' in \emph{Proc. INTERSPEECH}, 2016.

\bibitem{exp_tcd_3}
H.~Tak, J.~Patino, A.~Nautsch, N.~W.~D. Evans, and M.~Todisco, ``An explainability study of the constant q cepstral coefficient spoofing countermeasure for automatic speaker verification,'' in \emph{The Speaker and Language Recognition Workshop}, 2020.

\bibitem{graph2}
J.~weon Jung, H.-S. Heo, H.~Tak, H.~jin Shim, J.~S. Chung, B.-J. Lee, H.~jin Yu, and N.~W.~D. Evans, ``Aasist: Audio anti-spoofing using integrated spectro-temporal graph attention networks,'' \emph{IEEE International Conference on Acoustics, Speech and Signal Processing (ICASSP)}, pp. 6367--6371, 2022.

\bibitem{graph4}
F.~Chen, S.~Deng, T.~Zheng, Y.~He, and J.~Han, ``Graph-based spectro-temporal dependency modeling for anti-spoofing,'' in \emph{IEEE International Conference on Acoustics, Speech and Signal Processing (ICASSP)}, 2023, pp. 1--5.

\bibitem{headtoken}
Z.~Lu, H.~Xie, C.~Liu, and Y.~Zhang, ``Bridging the gap between vision transformers and convolutional neural networks on small datasets,'' in \emph{Advances in Neural Information Processing Systems}, A.~H. Oh, A.~Agarwal, D.~Belgrave, and K.~Cho, Eds., 2022.

\bibitem{xls_r}
A.~Babu, C.~Wang, A.~Tjandra, K.~Lakhotia, Q.~Xu, N.~Goyal, K.~Singh, P.~von Platen, Y.~Saraf, J.~M. Pino, A.~Baevski, A.~Conneau, and M.~Auli, ``{XLS-R}: Self-supervised cross-lingual speech representation learning at scale,'' in \emph{Interspeech}, 2021.

\bibitem{ssl_app_asr}
M.~Ravanelli, J.~Zhong, S.~Pascual, P.~Swietojanski, J.~Monteiro, J.~Trmal, and Y.~Bengio, ``Multi-task self-supervised learning for robust speech recognition,'' in \emph{IEEE International Conference on Acoustics, Speech and Signal Processing (ICASSP)}, 2020, pp. 6989--6993.

\bibitem{ssl_app_sv}
D.-T. Truong, R.~Tao, J.~Q. Yip, K.~A. Lee, and E.~S. Chng, ``Emphasized non-target speaker knowledge in knowledge distillation for automatic speaker verification,'' in \emph{IEEE International Conference on Acoustics, Speech and Signal Processing (ICASSP)}, 2024, pp. 10\,336--10\,340.

\bibitem{ssl_app_emotion}
E.~da~Silva~Morais, R.~Hoory, W.~Zhu, I.~Gat, M.~Damasceno, and H.~Aronowitz, ``Speech emotion recognition using self-supervised features,'' in \emph{IEEE International Conference on Acoustics, Speech and Signal Processing (ICASSP)}, 2022, pp. 6922--6926.

\bibitem{ssl_app_age_1}
D.-T. Truong, T.~T. Anh, and C.~E. Siong, ``Exploring speaker age estimation on different self-supervised learning models,'' in \emph{IEEE APSIPA ASC}, 2022, pp. 1950--1955.

\bibitem{ssl_app_age_2}
T.~Gupta, D.-T. Truong, T.~T. Anh, and C.~E. Siong, ``Estimation of speaker age and height from speech signal using bi-encoder transformer mixture model,'' in \emph{Proc. INTERSPEECH}, 2022, pp. 1978--1982.

\bibitem{graph3}
H.~Tak, M.~Todisco, X.~Wang, J.-w. Jung, J.~Yamagishi, and N.~Evans, ``Automatic speaker verification spoofing and deepfake detection using wav2vec 2.0 and data augmentation,'' in \emph{ODYSSEY 2022, The Speaker Language Recognition Workshop}, 2022.

\bibitem{resnet1}
H.~Tak, J.~Patino, M.~Todisco, A.~Nautsch, N.~W.~D. Evans, and A.~Larcher, ``End-to-end anti-spoofing with rawnet2,'' \emph{International Conference on Acoustics, Speech and Signal Processing (ICASSP)}, pp. 6369--6373, 2021.

\bibitem{asvspoof2019}
M.~Todisco, X.~Wang, V.~Vestman, M.~Sahidullah, H.~Delgado, A.~Nautsch, J.~Yamagishi, N.~W.~D. Evans, T.~H. Kinnunen, and K.~A. LEE, ``Asvspoof 2019: Future horizons in spoofed and fake audio detection,'' in \emph{Proc. INTERSPEECH}, 2019.

\bibitem{asvspoof2021}
X.~Liu, X.~Wang, M.~Sahidullah, J.~Patino, H.~Delgado, T.~Kinnunen, M.~Todisco, J.~Yamagishi, N.~Evans, A.~Nautsch, and K.~A. Lee, ``Asvspoof 2021: Towards spoofed and deepfake speech detection in the wild,'' \emph{IEEE/ACM Transactions on Audio, Speech, and Language Processing}, vol.~31, p. 2507–2522, 2023.

\bibitem{eer}
N.~Brümmer and E.~de~Villiers, ``The bosaris toolkit: Theory, algorithms and code for surviving the new dcf,'' 2013.

\bibitem{rawboost}
H.~Tak, M.~Kamble, J.~Patino, M.~Todisco, and N.~Evans, ``Rawboost: A raw data boosting and augmentation method applied to automatic speaker verification anti-spoofing,'' in \emph{IEEE International Conference on Acoustics, Speech and Signal Processing (ICASSP)}, 2022.

\end{thebibliography}

\end{document}